\documentclass[sigconf]{acmart}
\usepackage[utf8]{inputenc}

\AtBeginDocument{%
  \providecommand\BibTeX{{%
    \normalfont B\kern-0.5em{\scshape i\kern-0.25em b}\kern-0.8em\TeX}}}
\usepackage{booktabs}
\usepackage{multirow}
\usepackage{dsfont}
\usepackage{subcaption}

\copyrightyear{2020}
\acmYear{2020}
\setcopyright{acmcopyright}\acmConference[SIGIR '20]{Proceedings of the 43rd
International ACM SIGIR Conference on Research and Development in Information
Retrieval}{July 25--30, 2020}{Virtual Event, China}
\acmBooktitle{Proceedings of the 43rd International ACM SIGIR Conference on
Research and Development in Information Retrieval (SIGIR '20), July 25--30, 2020,
Virtual Event, China}
\acmPrice{15.00}
\acmDOI{10.1145/3397271.3401322}
\acmISBN{978-1-4503-8016-4/20/07}

\begin{document}
\fancyhead{}

\title{A Study of Neural Matching Models for Cross-lingual IR}

\author{Puxuan Yu and James Allan}
\affiliation{%
  \institution{Center for Intelligent Information Retrieval\protect\\
  University of Massachusetts Amherst}}
\email{{pxyu, allan}@cs.umass.edu}

\begin{abstract}

In this study, we investigate interaction-based neural matching models for ad-hoc cross-lingual information retrieval (CLIR) using cross-lingual word embeddings (CLWEs). With experiments conducted on the CLEF collection over four language pairs, we evaluate and provide insight into different neural model architectures, different ways to represent query-document interactions and word-pair similarity distributions in CLIR. This study paves the way for learning an end-to-end CLIR system using CLWEs.

\end{abstract}

\keywords{Cross-lingual Information Retrieval; Cross-lingual Word Embeddings; Neural Information Retrieval Model}

\maketitle

\section{Introduction}
\label{sec:intro}
CLIR is the task of retrieving documents in target language $L_t$ with queries written in source language $L_s$. The increasing popularity of projection-based weakly-supervised~\cite{xing2015normalized,glavavs2019properly,joulin2018loss} and unsupervised~\cite{conneau2017word,artetxe2018unsupervised} cross-lingual word embeddings has spurred unsupervised frameworks~\cite{litschko2018unsupervised} for CLIR, while in the realm of mono-lingual IR, interaction-based neural matching models~\cite{xiong2017end,guo2016deep,pang2016text} that utilize semantics contained in word embeddings have been the dominant force. This study fills the gap of utilizing CLWEs in neural IR models for CLIR. 

Traditional CLIR approaches translate either document or query using off-the-shelf SMT system such that query and document are in the same language. A number of researchers~\cite{ture2013flat,ture2014exploiting} later investigated utilizing translation table to build a probabilistic structured query~\cite{darwish2003probabilistic} in the target language. Recently, Litschko et al. showed that CLWEs are good translation resources by experimenting with a CLIR method (dubbed \textsf{TbT-QT}) that translates each query term in the source language to the nearest target language term in the CLWE space~\cite{litschko2018unsupervised}. CLWEs are obtained by aligning two separately trained embeddings for two languages in the same latent space, where a term in $L_s$ is proximate to its synonyms in $L_s$ and its translations in $L_t$, and vice versa. \textsf{TbT-QT} takes only the top-1 translation of a query term and uses the query likelihood model~\cite{ponte1998language} for retrieval. The overall retrieval performance can be damaged by vocabulary mismatch magnified with translation error. Using closeness measurement between query and document terms in the shared CLWE space as matching signal for relevance can alleviate the problem, but this area has not been extensively studied.

The reasons for the success of neural IR models for mono-lingual retrieval can be grouped into two categories:

\textbf{Pattern learning}: the construction of word-level query-document interactions enables learning of various matching patterns (e.g., proximity, paragraph match, exact match) via different neural network architectures.

\textbf{Representation learning}: models in which interaction features are built with differentiable operations (e.g., kernel pooling~\cite{xiong2017end}) allow customizing word embeddings via end-to-end learning from large-scale training data. 

Although representation learning is capable of further improving overall retrieval performance~\cite{xiong2017end}, it was shown in the same study that updating word embeddings requires large-scale training data to work well (more than 100k search sessions in their case). In CLIR, however, datasets usually have fewer than 200 queries per available language pair and can only support training neural models with smaller capacity. Therefore, we focus on the \textit{pattern learning} aspect of neural models.

In this study, we formulate the following research questions:

\begin{itemize}
    % \item \textbf{RQ1}: how is a neural model for CLIR different from mono-lingual IR?
    \item \textbf{RQ1}: how should a neural model for mono-lingual retrieval be adapted for CLIR?
    \item \textbf{RQ2}: how do neural models compare with each other and with unsupervised models for CLIR?
\end{itemize}

We answer these two main research questions with analysis (\S~\ref{sec:analysis}), experiments (\S~\ref{sec:experiment}) and discussions (\S~\ref{sec:discussion}) in the rest of the paper. 

\section{Analysis}\label{sec:analysis}

\subsection{Unsupervised CLIR Methods with CLWEs}

Two unsupervised CLIR approaches using CLWEs are proposed by Litschko et al.~\cite{litschko2018unsupervised}.  \textbf{BWE-Agg} ranks documents with respect to a query using the cosine similarity of query and document embeddings, obtained by aggregating the CLWEs of their constituent terms. The simpler version, namely \textsf{BWE-Agg-Add}, takes the average embeddings of all terms for queries and documents, while the more advanced version \textsf{BWE-Agg-IDF} builds document embeddings by weighting terms with their inverse document frequencies. \textbf{TbT-QT}, as described in \S~\ref{sec:intro}, first translates each query term to its nearest cross-lingual neighbor term and then adopts query-likelihood in mono-lingual setting. These two approaches represent different perspectives towards CLIR using CLWEs. \textsf{BWE-Agg} builds query and document representations out of CLWEs but completely neglects exact matching signals, which play important roles in IR. Also, although query and document terms are weighted based on IDF, using only one representation for a long document can fail to emphasize the section of a document that is truly relevant to the query. \textsf{TbT-QT} only uses CLWEs as query translation resources and adopts exact matching in a mono-lingual setting, so its performance is heavily dependent on the translation accuracy (precision@1) of CLWEs. Analytically, an interaction-based neural matching model that starts with word level query-document interactions and considers both exact and similar matching can make up for the shortcomings of the above two methods. 

\vspace{-1ex}

\subsection{Neural IR Models}

\subsubsection{Background}

For interaction-based matching models, we select three representative models (\textsf{MatchPyramid}~\cite{pang2016text,pang2016study}, \textsf{DRMM}~\cite{guo2016deep} and \textsf{KNRM}~\cite{xiong2017end}) from the literature for analysis and experiments.

\textbf{MatchPyramid}: The \textsf{MatchPyramid}~\cite{pang2016text,pang2016study} (\textsf{MP} for short) is one of the earliest models that starts with capturing word-level matching patterns for retrieval. It casts the ad-hoc retrieval task as a series of image recognition problems, where the ``image'' is the matching matrix of a query-document pair $(q,d)$, and each ``pixel'' is the interaction value of a query term $q_i$ and a document term $d_j$. Typical interaction functions are cosine similarity, dot product, Gaussian kernel, and indicator function (for exact match). The intuition behind hierarchical convolutions and pooling is to model phrase, sentence and even paragraph level matching patterns.

\textbf{DRMM}: The \textsf{DRMM}~\cite{guo2016deep} model uses a matching histogram to capture the interactions of a query term with the whole document. The valid interval of cosine similarity (i.e., $[-1,1]$) is discretized into a fixed number of bins such that a matching histogram is essentially a fixed-length integer vector. Features from different histograms are weighted based on attention calculated on query terms. \textsf{DRMM} is not position-preserving, as the authors claim that relevance matching is not related to term order. 

\textbf{K-NRM}: The \textsf{KNRM}~\cite{xiong2017end} model takes matrix representation for query-document interaction (similar to \textsf{MP}), but ``categorizes'' interactions into different levels of cosine similarities (similar to \textsf{DRMM}), using Gaussian kernels with different mean value $\mu$. The distinct advantage of \textsf{KNRM} over \textsf{DRMM} is that the former allows gradient to pass through Gaussian kernels, and therefore supports end-to-end learning of embeddings. 

\subsubsection{Mono-lingual to Cross-lingual}

According to results reported in respective studies~\cite{pang2016text,guo2016deep,xiong2017end}, the relative performance of three models for mono-lingual IR should be \textsf{KNRM} $>$ \textsf{DRMM} $>$ \textsf{MP}, even when embedding learning is turned off with \textsf{KNRM}. Tweaking a neural model for support of CLIR is trivial: instead of considering interaction value as two terms' similarity in a mono-lingual embedded space, we consider the proximity of their representations in the shared cross-lingual embedded space. However, there are several matters to consider while making the transition:

\textbf{Exact matching signals}: The significant difference between cross-lingual and mono-lingual IR is that the former (almost) never encounters exact match of terms in different languages. However, neglecting such factors can be costly for models like \textsf{MP}, the disadvantage of which when compared to the other two models is the inability to capture exact and similarity matching signals at the same time. To this end, we first define in CLIR the exact matching of two terms (in different languages) as their cosine similarity in the CLWE space exceeding a certain threshold value $\eta$. 
We then implement a hybrid version, namely \textsf{MP-Hybrid}, that joins exact and soft matching signals extracted from interaction matrices built with indicator function and cosine similarity function, such that ranking features from dual channels are concatenated for an MLP to predict a ranking score.

\textbf{Word-pair similarity distribution}: The cosine similarities of two terms with close meanings but in different languages are distributed differently than those in the same language. 
Specifically, the top word-pair similarity distributions of CLWEs tend to have smaller mean and variance.
In an example shown in Table~\ref{tab:example}, the cosine similarity of the five closest words to ``telephone'' in English embedded space\footnote{\url{https://dl.fbaipublicfiles.com/fasttext/vectors-english/wiki-news-300d-1M.vec.zip}} ranges from 0.818 to 0.669, while in aligned English-Spanish embedded space\footnote{\url{https://dl.fbaipublicfiles.com/fasttext/vectors-aligned/wiki.es.align.vec}}, it ranges from 0.535 to 0.520. The similarity distribution affects histogram construction of \textsf{DRMM} and similarly for the kernel pooling of \textsf{KNRM}. The distribution also affects the exact matching threshold value $\eta$ for related variants of \textsf{MP}. Since the cosine similarity of a query term and its perfectly correct translation can be less than 0.6, setting $\eta$ too high can lead to failure of capturing positive matching signals.

\protect
\begin{table}[t]

\caption{Cosine similarities of the top-5 closest words to ``telephone'' in an English embedding space (EN) and in an aligned English-Spanish embedding space (ES).}
\vspace{-2ex}
\label{tab:example}
\resizebox{\linewidth}{!}{%
\begin{tabular}{|c|c|c|c|c|c|}
\hline
\multirow{2}{*}{EN} & phone       & telephones & Telephone  & landline  & rotary-dial \\ \cline{2-6} 
                    & 0.818       & 0.761      & 0.720      & 0.694     & 0.669       \\ \hline
\multirow{2}{*}{ES} & telefónicos & teléfono   & telefónica & telefónia & telefóno    \\ \cline{2-6} 
                    & 0.535       & 0.522      & 0.522      & 0.520     & 0.520       \\ \hline
\end{tabular}%
}

\vspace{-4ex}

\end{table}

\textbf{Vocabulary mismatch and translation error}: Query translation based CLIR methods (e.g., \textsf{TbT-QT}~\cite{litschko2018unsupervised}) first translate queries from $L_s$ to $L_t$, then use mono-lingual retrieval in $L_t$. Apart from the inherent vocabulary mismatch problem within $L_t$, the translation error from $L_s$ to $L_t$ has to be also counted. Looking at the example in Table~\ref{tab:example}, \textsf{TbT-QT} would look for occurrence of ``telefónicos'' in the collection, and documents containing only the correct translation (``teléfono'') would be overlooked. Interaction-based neural matching models alleviate this issue by giving partial credit to sub-optimal nearest neighbors, which in many cases are the correct translations. To demonstrate the necessity of directly using cross-lingual word embedding similarity as interaction for neural models, we conduct comparative experiments where queries are first translated term-by-term like \textsf{TbT-QT} using CLWEs, then used for retrieval in mono-lingual setting. Such models are referred to as \textsf{\{MP,DRMM,K-NRM\}-TbT-QT}, respectively.

\section{Experiments}\label{sec:experiment}

\noindent

\textbf{Datasets}: We evaluate the models on the CLEF test suite for the CLEF 2000-2003 campaigns. We select four language pairs: English (EN) queries to \{Dutch (NL), Italian (IT), Finnish (FI), Spanish (ES)\} documents. All documents for the four languages are used for evaluation, and are truncated to preserve the first 500 tokens for computational efficiency~\cite{pang2016study}. The statistics of the evaluation datasets are shown in Table~\ref{tab:stats}. The titles of CLEF topics are used as English queries. All queries and documents are lower-cased, with stopwords, punctuation marks and one-character tokens removed. 

\textbf{Cross-lingual word embeddings}: We adopt the pre-aligned fastText CLWEs\footnote{\url{https://fasttext.cc/docs/en/aligned-vectors.html}}. Mono-lingual fastText embeddings are trained on Wikipedia corpus in respective languages, and aligned using weak supervision from a small bilingual lexicon with the RCSLS loss as the optimization objective~\cite{joulin2018loss}.

\begin{table}[t]
 \caption{Basic statistics of CLEF data for evaluation: number of queries (\#queries), number of documents (\#docs), average number of relevant documents per query (\#rel), and average number of labeled documents per query (\#label).}
 \vspace{-2ex}
 \label{tab:stats}
 \centering
 \begin{tabular}{l|r|r|r|r} 
  Lang. Pair & EN$\to$ NL & EN$\to$ IT  & EN$\to$ FI & EN$\to$ ES   \\ [0.5ex]
  \toprule
 \#queries & 160 & 160 & 90 & 160\\
 \#docs & 42,734 & 40,320 & 16,351 & 46,540 \\
 \#rel & 29.1 & 19.5 & 10.9 & 49.5 \\
 \#label & 375.4 & 338.3 & 282.6 & 372.7 \\
 \end{tabular}
 \vspace{-2ex}
\end{table}

\textbf{Model specifications}: We implemented two CLWEs based unsupervised CLIR algorithms \textsf{BWE-Agg} and \textsf{TbT-QT} as baselines~\cite{litschko2018unsupervised}. In addition to the query likelihood model in the original study, we pair \textsf{TbT-QT} with \textsf{BM25} to investigate the influence of retrieval models to queries translated using CLWEs. 

We experiment with five variants of the \textsf{MP} model, two for the \textsf{DRMM} model and two for the \textsf{KNRM} model. As the interaction value of query term $q_i$ and document term $d_j$, \textsf{\{MP,DRMM,KNRM\}-Cosine} uses the cosine similarity $\cos{(q_i,d_j)}=\vec{q_i}^\intercal\vec{d_j}/(||\vec{q_i}||\cdot ||\vec{d_j}||)$, \textsf{MP-Gaussian} uses $e^{-||\vec{q_i}-\vec{d_j}||^2}$, and \textsf{MP-Exact} takes $\mathds{1}_{\{\cos{(q_i,d_j)}\ge \eta\}}$, where $\eta$ is a pre-defined threshold value (set to 0.3 for Table~\ref{tab:results}). \textsf{MP-Hybrid} concatenates the flattened features after dynamic pooling layer from \textsf{MP-Cosine} and \textsf{MP-Exact} into one vector, and uses an MLP to predict a final score. \textsf{\{MP,DRMM,KNRM\}-TbT-QT} is equal to first translating query $q$ to target language query $\text{tr}(q)$, and running $\text{tr}(q)$ with \textsf{\{MP,DRMM,KNRM\}-Cosine} model.

For the \textsf{MP} model, we adopt one layer convolution with kernel size set to $3\times3$, dynamic pooling size set to $5\times1$, and kernel count set to 64. For the \textsf{DRMM} model, we adopt the log-count-based histogram with bin size set to 30. For the \textsf{KNRM} model, kernel count is set to 20 and standard deviation of each Gaussian kernel is set to 0.1. All decisions made above are based on extensive hyper-parameter tuning that first prioritizes generalizable retrieval performance then computational efficiency and model simplicity. 

\textbf{Model training}: All neural models in the experiments are trained with the pairwise hinge loss. Given a triple $(q,d+,d-)$, where document $d+$ is relevant and document $d-$ is non-relevant with respect to query $q$, the loss function is defined as:
$$L(q,d+,d-;\Theta)=\max\{0,1-s(q,d+)+s(q,d-)\}$$
\noindent
where $s(q,d)$ denotes the predicted matching score for $(q, d)$, and $\Theta$ represents the learnable parameters in the neural network. Note that we randomly select documents that are explicitly labeled non-relevant (-1) as negative samples for training. Five negative $(q,d)$ pair are sampled for each positive pair. We apply stochastic gradient descent method Adam~\cite{kingma2014adam} (learning rate=1e-3) in mini-batches (64 in size) for optimization. The maximum number of training epochs allowed is 20. 

\textbf{Evaluation}: As the CLEF dataset uses binary relevance judgement, we adopt MAP as the evaluation metric. In order to conduct evaluation on enough queries that conclusions can possibly be statistically significant, we adopt 5-fold cross-validation with validation and test sets. Statistical significant tests are performed using the two-tailed paired t-test at the $0.05$ level.

\begin{table}[t]
  \caption{MAP performance of all CLIR methods. \textbf{Boldfaced} is the best performer in each language pair. Underlined is the best \textsf{MP} variant.} 
  \vspace{-2ex}
  \label{tab:results}
  \begin{tabular}{l|c|c|c|c}
  Lang. Pair & EN$\to$ NL & EN$\to$ IT  & EN$\to$ FI & EN$\to$ ES   \\
    \toprule
    \textsf{BWE-Agg-Add} & .237 & .173 & .170 & .297 \\
    \textsf{BWE-Agg-IDF} & .246 & .178 & .180 & .298\\
    \midrule
    \textsf{TbT-QT-BM25} & .240 & .231 & .122 & .341\\
    \textsf{TbT-QT-QL} & .297 & .268 & .126 & .387\\
    \midrule
    \textsf{MP-Cosine} & \underline{.348} & \underline{.331} & \underline{.254} & .423\\
    \textsf{MP-Gaussian} & .322 & .319 & .203 & .405\\
    \textsf{MP-Exact} & .327 & .295 & .202 & .415\\
    \textsf{MP-Hybrid} & .343 & .326 & .243 & \underline{.427} \\
    \textsf{MP-TbT-QT} & .327 & .300 & .195 & .409\\
    \midrule
    \textsf{DRMM-Cosine} & \textbf{.374} & \textbf{.352} & \textbf{.304} & \textbf{.462}\\
    \textsf{DRMM-TbT-QT} & .345 & .324 & .193 & .450\\
    \midrule
    \textsf{KNRM-Cosine} & .368 & .313 & .286 & .423\\
    \textsf{KNRM-TbT-QT} & .329 & .288 & .200 & .405\\
    \bottomrule
  \end{tabular}
  \vspace{-2ex}
\end{table}

\section{Discussion and Conclusion}
\label{sec:discussion}

\subsection{Parsing Results}

The experimental results of CLIR on four language pairs are reported in Table~\ref{tab:results}.  \textsf{TbT-QT} generally works better than \textsf{BWE-Agg} except for EN$\to$FI. This might indicate that the English-Finnish CLWEs are not aligned well to provide quality top-1 query term translation. The larger gaps between \textsf{\{MP,DRMM,K-NRM\}-Cosine} and \textsf{\{MP,DRMM,K-NRM\}-TbT-QT} for EN-FI than the other three language pairs reinforce this argument. All neural models achieve \textit{statistically significant} improvement over heuristic baselines. \textsf{DRMM-Cosine} consistently achieves the best performance for all language pairs. Although \textsf{DRMM} and \textsf{KNRM} are conceptually similar, the former performs significantly better, with \textsf{KNRM}'s embedding layer kept frozen. The attention mechanism applied to query terms for \textsf{DRMM} can be a factor. On EN$\to$\{IT,ES\}, the \textsf{MP} model performs on par with or better than \textsf{KNRM}. This finding indicates that the convolution plus dynamic pooling architecture can also be an option for learning an end-to-end CLIR model. Comparing different approaches to build query-document interaction matrices for \textsf{MP}, it is clear that cosine similarity of source language query term and target language document term in the CLWE space is the best choice, which contradicts the conclusions in the study of mono-lingual IR~\cite{pang2016study} where Gaussian kernel and indicator function are found to work better. The exact matching variant \textsf{MP-Exact} we proposed works reasonably well, indicating that most decisions of relevance are influenced by top similarity matching signals. The hybrid variant \textsf{MP-Hybrid} we propose improves upon \textsf{MP-Exact} but does not outperform \textsf{MP-Cosine} except for EN$\to$ES. This is expected because matching signals from \textsf{MP-Exact} are not from truly exact matches of terms, but are derived from cosine similarity matrices as in \textsf{MP-Cosine}. The combination of two models results in redundant information. The fact that \textsf{\{MP,DRMM,K-NRM\}-TbT-QT} outperform baseline approaches but are not as good as respective cosine variants demonstrates (1)~the effectiveness of pattern learning of neural models; and (2)~the necessity to directly build cross-lingual interactions of query and document in two languages, rather than building interactions after translation.

\subsection{Word-pair Similarity Distribution}

\begin{figure}[t]
    \begin{subfigure}[t]{0.85\linewidth}   
        \includegraphics[width=1\linewidth]{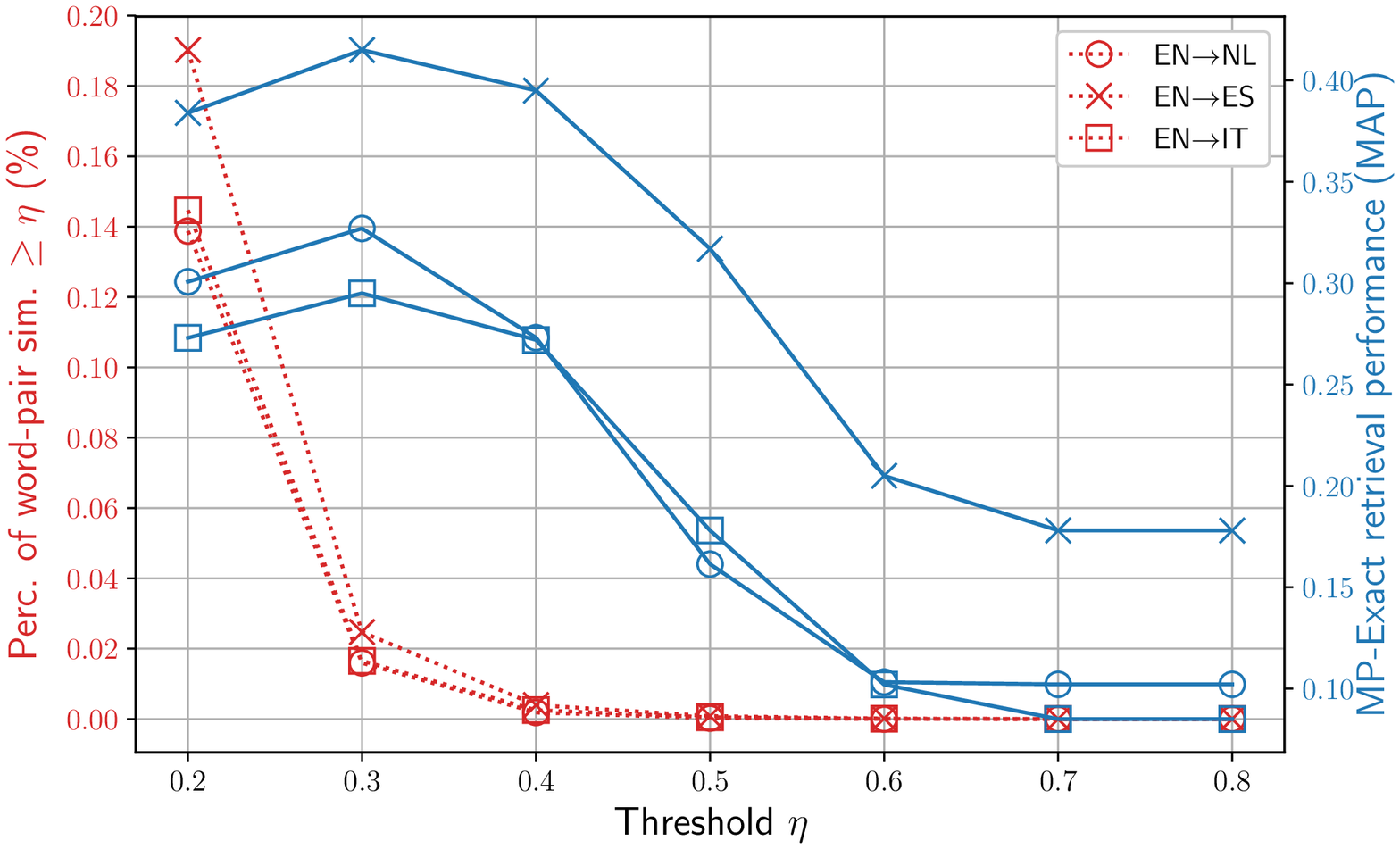}
        \caption{EN$\to$\{NL,ES,IT\}}
        \label{fig:dist1}
    \end{subfigure}
    \begin{subfigure}[t]{0.48\linewidth}
        \includegraphics[width=1\linewidth]{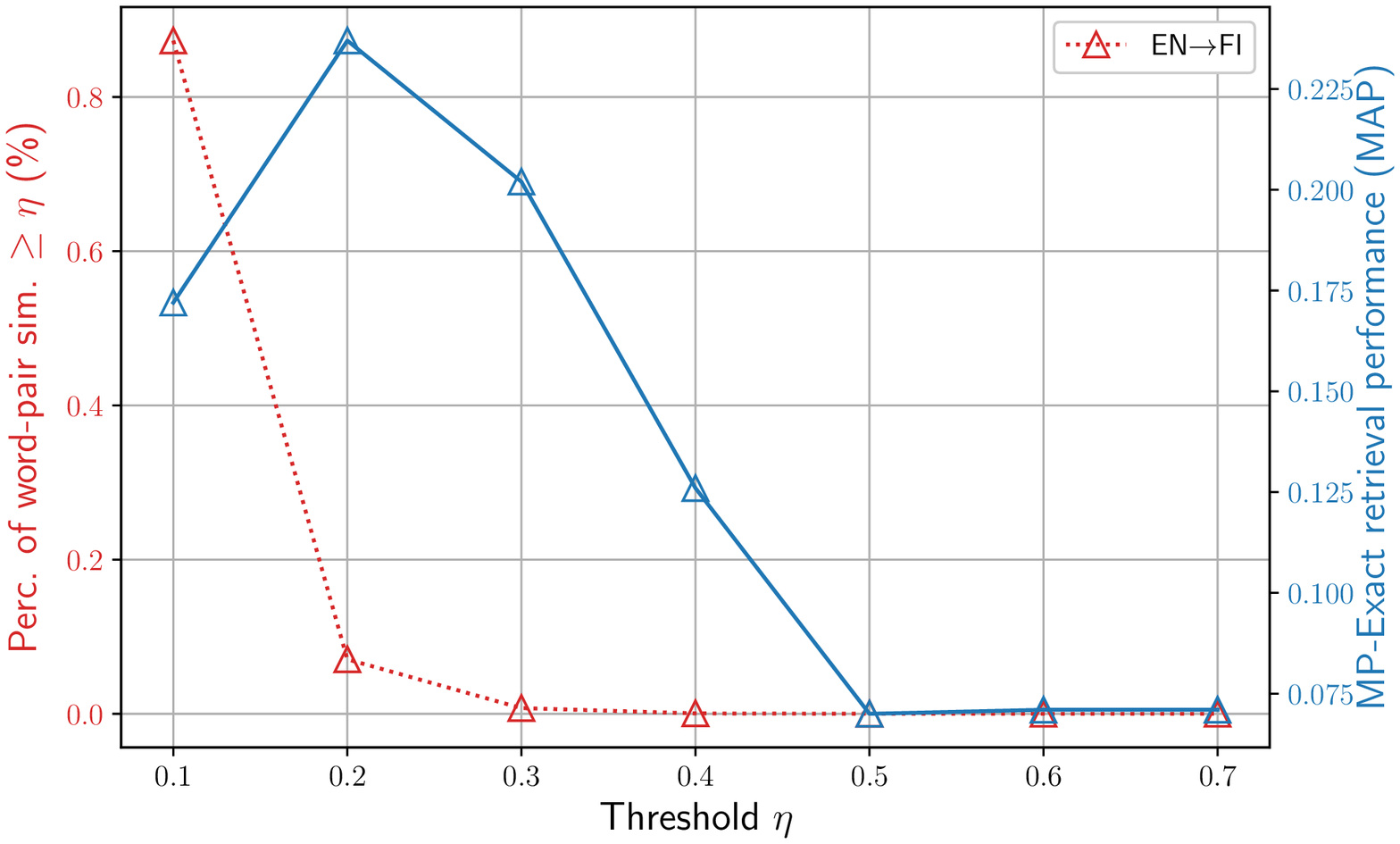}
        \caption{EN$\to$FI}
        \label{fig:dist2}
    \end{subfigure}
    \begin{subfigure}[t]{0.48\linewidth}
        \includegraphics[width=1\linewidth]{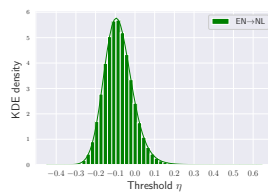}
        \caption{EN$\to$NL}
        \label{fig:dist3}
    \end{subfigure}
\caption{(a,b) -- Red: percentage of cross-lingual word pair with similarity $\ge \eta$; Blue:  \textsf{MP-Exact} retrieval performance with different similarity threshold value $\eta$. (c): Similarity distribution of word-pairs in the EN$\to$NL collection.}
\vspace{-4ex}
\label{fig:dist}
\end{figure}

The distribution of word pair similarities influences the exact matching threshold $\eta$ in \textsf{MP-Exact}, the query translation strategy in \textsf{TbT-QT}, and the embedding fine-tuning for an end-to-end model. We take source language terms in the queries and target language terms in the documents, calculate their pairwise cosine similarities in the aligned CLWE space, and plot the similarity distributions. In Figure~\ref{fig:dist1} and~\ref{fig:dist2}, we show in red the percentage of cross-lingual word-pairs with similarity above $\eta$. The three distributions in Figure~\ref{fig:dist1} are very similar at tail ($\eta\ge0.2$), therefore the corresponding \textsf{MP-Exact}'s performance peaks at the same $\eta=0.3$. EN$\to$FI is distributed differently but the pattern shown is similar (Figure~\ref{fig:dist2}). The shapes of cross-lingual similarity distribution for all four language pairs are very similar, therefore we only plot EN$\to$NL in Figure~\ref{fig:dist3} for demonstration. Mono-lingual similarity distribution in Xiong et al.'s study~\cite{xiong2017end} has large variance, positive mean, strong positive skewness and high density at large $\eta$. In comparison, the cross-lingual similarity distribution (Figure~\ref{fig:dist3}) has small variance, negative mean, no obvious skewness to the left or right, and the density drops low and flat after $\eta=0.4$, where word-pairs are considered highly similar (i.e., quality translations). This provides insights into why top-1 translation with CLWEs is not necessarily significantly better than translations ranked at slightly lower positions.

\vspace{-2ex}

\subsection{Conclusions}

\noindent

\textit{Answer to RQ1}: To adapt a neural model for CLIR, exact matching representations, cross-lingual word-pair similarity distribution, and translation error using CLWEs have to be considered. In specific model settings, choices of interaction representations and hyper-parameters (e.g., dynamic pooling size at document side for \textsf{MP}) are found to be different from mono-lingual IR.

\textit{Answer to RQ2}: Neural matching models experimented in this study all outperform baselines using CLWEs. The \textsf{DRMM} achieves the best results across the board, while \textsf{MP} and \textsf{KNRM} perform inconsistently on different language pairs.

Moving forward, a worthwhile endeavor will be to investigate an end-to-end neural model that learns from large-scale CLIR data.

\subsection*{Acknowledgements}

This work was supported in part by the Center for Intelligent Information Retrieval and in part by the Air Force Research Laboratory (AFRL) and IARPA under contract \#FA8650-17-C-9118 under subcontract \#14775 from Raytheon BBN Technologies Corporation. Any opinions, findings and conclusions or recommendations expressed in this material are those of the authors and do not necessarily reflect those of the sponsor.

\bibliographystyle{ACM-Reference-Format}
\bibliography{sample-bibliography.bib}

\end{document}